\documentclass[twocolumn]{jpsj3}

\usepackage{color}
\usepackage{bm}

\title{%
Spontaneous Charge Current in a Doped Weyl Semimetal
}

\author{%
Yositake Takane
}

\inst{%
Department of Quantum Matter, Graduate School of Advanced Sciences of Matter,\\
Hiroshima University, Higashihiroshima, Hiroshima 739-8530, Japan
}

\recdate{ \hspace{50mm} }

\abst{%
A Weyl semimetal hosts low-energy chiral surface states, which appear
to connect a pair of Weyl nodes in reciprocal space.
As these chiral surface states propagate in a given direction,
a spontaneous circulating current is expected to appear
near the surface of a singly connected Weyl semimetal.
This possibility is examined
by using a simple model with particle-hole symmetry.
It is shown that no spontaneous charge current appears
when the Fermi level is located at the band center.
However, once the Fermi level deviates from the band center,
a spontaneous charge current appears to circulate around
the surface of the system and its direction of flow
is opposite for the cases of electron doping and hole doping.
These features are qualitatively unchanged 
even in the absence of particle-hole symmetry.
The circulating charge current is shown to be robust against weak disorder.
}

\begin{document}
\sloppy
\maketitle

\section{Introduction}

A Weyl semimetal possesses a pair of, or pairs of, nondegenerate Dirac cones
with opposite chirality.~\cite{shindo,murakami,wan,yang,burkov1,burkov2,WK,
delplace,halasz,sekine}
The pair of Dirac cones can be nondegenerate
if time-reversal symmetry or inversion symmetry is broken.
The electronic property of a Weyl semimetal is significantly influenced by
the position of a pair of Weyl nodes in reciprocal and energy spaces,
where a Weyl node represents the band-touching point of each Dirac cone.
In the absence of time-reversal symmetry,
a pair of Weyl nodes is separated in reciprocal space.
In this case, low-energy states with chirality appear on the surface of
a Weyl semimetal~\cite{wan} if the Weyl nodes are projected onto
two different points in the corresponding surface Brillouin zone.
A notable feature of such chiral surface states is that they propagate
only in a given direction, which depends on the position of the Weyl nodes.
This gives rise to an anomalous Hall effect.~\cite{burkov1}
If inversion symmetry is also broken in addition to time-reversal symmetry,
a pair of Weyl nodes is also separated in energy space.
In this case, the propagating direction of chiral surface states
is tilted according to the deviation of the Weyl nodes.
If only inversion symmetry is broken, no chiral surface state appears
since the pair of Weyl nodes coincides in reciprocal space.
To date, some materials have been experimentally identified
as Weyl semimetals.~\cite{weng,huang1,xu1,lv1,lv2,xu2,souma,kuroda}

Let us focus on the case in which a Weyl semimetal with a pair of Weyl nodes
at $\mib{k}_{\pm} = (0,0,\pm k_{0})$ is
in the shape of a long prism parallel to the $z$ axis.
In this case, chiral surface states appear on the side of the system.
In the presence of inversion symmetry, they typically propagate in a direction
perpendicular to the $z$ axis; thus, we expect that a spontaneous charge
current appears to circulate in the system near the side surface.
If inversion symmetry is additionally broken, the propagating direction is
tilted to the $z$ direction; thus, an electron in the chiral surface state
shows spiral motion around the system.~\cite{baireuther}
Thus, we expect that a spontaneous charge current has
a nonzero component in the $z$ direction.
This longitudinal component must be canceled out
by the contribution from bulk states
if they are integrated over a cross section parallel to the $xy$ plane.

In this paper, we theoretically examine whether a spontaneous charge current
appears in the ground state of a Weyl semimetal.
Our attention is focused on the case where time-reversal symmetry is broken.
We calculate the spontaneous charge current induced near the side
of the system by using a simple model with particle-hole symmetry.
We find that no spontaneous charge current appears when the Fermi level,
$E_{F}$, is located at the band center, which is set equal to $0$ hereafter,
implying that the contribution from chiral surface states
is completely canceled out by that from bulk states.
However, once $E_{F}$ deviates from the band center,
the spontaneous charge current appears to circulate around the side surface
of the system and its direction of flow is opposite for the cases of
electron doping (i.e., $E_{F} > 0$) and hole doping (i.e., $E_{F} < 0$).
The circulating charge current is shown to be robust against weak disorder.
In the absence of inversion symmetry,
we show that chiral surface states induce the longitudinal component of
a spontaneous charge current near the side surface, which is compensated by
the contribution from bulk states appearing beneath the side surface.
This longitudinal component is shown to be fragile against disorder.

In the next section, we present a tight-binding model for Weyl semimetals
and show the absence of a spontaneous charge current when the Fermi level
is located at the band center.
In Sect.~3, we derive a tractable continuum model from the tight-binding model
and analytically determine the magnitude of a spontaneous charge current
induced by the deviation of the Fermi level from the band center.
In Sect.~4, we numerically study the behaviors of
a spontaneous charge current by using the tight-binding model.
We also examine the effect of disorder on the spontaneous charge current.
The last section is devoted to a summary and discussion.
We set $\hbar = 1$ throughout this paper.

\section{Model}

Let us introduce a tight-binding model for Weyl semimetals
on a cubic lattice with lattice constant $a$.
Its Hamiltonian is given by
$H = H_{0}+H_{x}+H_{y}+H_{z}$ with~\cite{yang,burkov1}
\begin{align}
   H_{0}
 & = \sum_{l,m,n} |l,m,n \rangle h_{0} \langle l,m,n| ,
         \\
   H_{x}
 & = \sum_{l,m,n}
     \left\{ |l+1,m,n \rangle h_{x} \langle l,m,n|
             + {\rm h.c.} \right\} ,
         \\
   H_{y}
 & = \sum_{l,m,n}
     \left\{ |l,m+1,n \rangle h_{y} \langle l,m,n|
             + {\rm h.c.} \right\} ,
         \\
   H_{z}
 & = \sum_{l,m,n}
     \left\{ |l,m,n+1 \rangle h_{z} \langle l,m,n|
             + {\rm h.c.} \right\} ,
\end{align}
where the indices $l$, $m$, and $n$ are respectively used to specify
lattice sites in the $x$, $y$, and $z$ directions and
\begin{align}
  |l,m,n \rangle
  \equiv  \left[ |l,m,n \rangle_{\uparrow}, |l,m,n \rangle_{\downarrow}
     \right] 
\end{align}
represents the two-component state vector with
$\uparrow, \downarrow$ corresponding to the spin degree of freedom.
The $2 \times 2$ matrices are
\begin{align}
   h_{0}
 & = \left[ 
       \begin{array}{cc}
         2t\cos(k_{0}a) + 4B & 0 \\
         0 & -2t\cos(k_{0}a) - 4B
       \end{array}
     \right] ,
               \\
   h_{x}
 & = \left[ 
       \begin{array}{cc}
         -B & \frac{i}{2}A \\
         \frac{i}{2}A & B
       \end{array}
     \right] ,
               \\
   h_{y}
 & = \left[ 
       \begin{array}{cc}
         -B & \frac{1}{2}A \\
         -\frac{1}{2}A & B
       \end{array}
     \right] ,
               \\
   h_{z}
 & = \left[ 
       \begin{array}{cc}
         -t+i\gamma & 0 \\
         0 & t+i\gamma
       \end{array}
     \right] ,
\end{align}
where $0 < k_{0} < \pi/a$, and the other parameters, $A$, $B$, $t$,
and $\gamma$, are assumed to be real and positive.
The Fourier transform of $H$ is expressed as
\begin{align}
  \mathcal{H}(\mib{k}) =
     \left[ 
        \begin{array}{cc}
          \Lambda(\mib{k})+2\gamma\sin(k_{z}a) & \Theta_{-}(k_{x},k_{y}) \\
          \Theta_{+}(k_{x},k_{y}) & -\Lambda(\mib{k})+2\gamma\sin(k_{z}a)
        \end{array}
     \right] ,
\end{align}
where $\Lambda(\mib{k}) = \Delta(k_{z})
+2B\sum_{\alpha=x,y}[1-\cos(k_{\alpha}a)]$
and $\Theta_{\pm}(k_{x},k_{y}) = A[\sin(k_{x}a)\pm i\sin(k_{y}a)]$ with
\begin{align}
  \Delta(k_{z}) = - 2t\left[\cos(k_{z}a)-\cos(k_{0}a)\right] .
\end{align}
It can be seen that inversion symmetry is broken if $\gamma \neq 0$.
The energy dispersion of this model is given as
\begin{align}
           \label{eq:exp-E}
   E =
 & 2\gamma\sin(k_{z}a)
   \pm \Big\{  A^{2}\bigl(\sin^{2}(k_{x}a)+\sin^{2}(k_{y}a)\bigr)
                 \nonumber \\
 & \hspace{-4mm}
             + \left[\Delta(k_{z})
                  +2B\bigl(2-\cos(k_{x}a)-\cos(k_{y}a)\bigr)\right]^{2}
       \Big\}^{\frac{1}{2}} ,
\end{align}
indicating that a pair of Weyl nodes appears
at $\mib{k}_{\pm} = (0,0,\pm k_{0})$.
Note that $E = 0$ at the band center.
The energy of each Weyl node is located at the band center of $E = 0$
at $\gamma = 0$, whereas it deviates from there if $\gamma \neq 0$.
In Sect.~4, we focus on the lattice system of a rectangular parallelepiped
with $L$ sites in both the $x$ and $y$ directions
and $N$ sites in the $z$ direction (see Fig.~1).
In this setup, chiral surface states appear on its side.
%%%%%%%%%%%%%%%%%%
\begin{figure}[btp]
\begin{center}
\includegraphics[width=5cm]{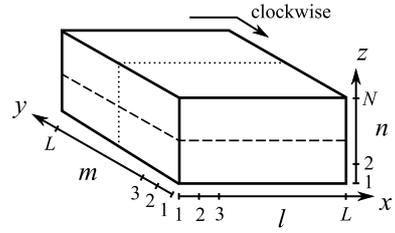}
\end{center}
\caption{
Lattice system considered in the text: a rectangular parallelepiped
with $L$ sites in both the $x$ and $y$ directions
and $N$ sites in the $z$ direction.
}
\end{figure}
%%%%%%%%%%%%%%%%%%

Now, we show that no spontaneous charge current appears
if $E_{F}$ is located at the band center on the basis of
the particle-hole symmetry inherent in the model introduced above.
Let us define the operator $\Gamma_{\rm ph}$ as
\begin{align}
  \Gamma_{\rm ph} = \sigma_{x}K ,
\end{align}
where $\sigma_{x}$ and $K$ are respectively the $x$ component of
the Pauli matrices and the complex conjugate operator.
The tight-binding Hamiltonian $H$ satisfies
\begin{align}
    \label{eq:ph-symmet}
  \Gamma_{\rm ph}^{-1}H\Gamma_{\rm ph} = - H .
\end{align}
Let us denote eigenstates of $H$ in the valence band as $|q\rangle_{v}$
and those in the conduction band as $|q\rangle_{c}$
with $q = 1,2,3,\dots$,
where $H|q\rangle_{v}=\epsilon_{q}^{v}|q\rangle_{v}$
with $\epsilon_{q}^{v} < 0$ and $H|q\rangle_{c}=\epsilon_{q}^{c}|q\rangle_{c}$
with $\epsilon_{q}^{c} > 0$.
Here, $q$ labels the eigenstates in descending order
(i.e., $0\ge\epsilon_{1}^{v}\ge\epsilon_{2}^{v}\ge\epsilon_{3}^{v}\ge\dots$)
in the valence band and in ascending order
(i.e., $0\le\epsilon_{1}^{c}\le\epsilon_{2}^{c}\le\epsilon_{3}^{c}\le\dots$)
in the conduction band.
Equation~(\ref{eq:ph-symmet}) allows us to set
$\epsilon_{q}^{v}=-\epsilon_{q}^{c}$ with
\begin{align}
  |q\rangle_{c} = \Gamma_{\rm ph} |q\rangle_{v} .
\end{align}

We introduce the charge current operator $j_{\alpha}$ defined on
an arbitrary site, where $\alpha = x,y,z$ specifies the direction of flow.
For example, the current operator $j_{x}$ on the $(l,m,n)$th site is given by
\begin{align}
  j_{x}
   = -e(-i)\bigl[ |l+1,m,n\rangle h_{x} \langle l,m,n| - {\rm h.c.}
            \bigr] .
\end{align}
It may be more appropriate to state that this is defined on the link
connecting the $(l,m,n)$th and $(l+1,m,n)$th sites.
We can show that any $j_{\alpha}$ is invariant
under the transformation of $\Gamma_{\rm ph}$ as
\begin{align}
  \Gamma_{\rm ph}^{-1}j_{\alpha}\Gamma_{\rm ph} = j_{\alpha} .
\end{align}
In the ground state with $E_{F} = 0$,
the expectation value of any $j_{\alpha}$ is expressed as
\begin{align}
  \langle j_{\alpha} \rangle
 = \sum_{q \ge 1} {}_{v}\langle q |j_{\alpha}|q\rangle_{v} .
\end{align}
By using the relations given above and the completeness of the set of
eigenstates consisting of $\{|q\rangle_{v}\}$ and $\{|q\rangle_{c}\}$,
we can show that
\begin{align}
  \langle j_{\alpha} \rangle
 & = \frac{1}{2} \sum_{q \ge 1}
     \left[ {}_{v}\langle q |j_{\alpha}|q\rangle_{v}
           +{}_{c}\langle q |j_{\alpha}|q\rangle_{c}
     \right]
       \nonumber \\
 & = \frac{1}{2}{\rm tr}\{ j_{\alpha} \} = 0 ,
\end{align}
indicating that a spontaneous charge current completely vanishes
everywhere in the system at $E_{F} = 0$.
That is, although chiral surface states carry
a circulating charge current, their contribution is completely canceled out
by that from bulk states.

Note that the above argument based on particle-hole symmetry
is not restricted to the two-orbital model used in this study
and is also applicable to the four-orbital model
introduced in Ref.~\citen{vazifeh}.

\section{Analytical Approach}

Before performing numerical simulations in the case of $E_{F} \neq 0$,
we analytically study the behaviors of chiral surface states
in a cylindrical Weyl semimetal.
To do so, we apply the analytical approach given in Ref.~\citen{takane1},
which was developed to describe unusual electron states in a Weyl semimetal:
chiral surface states~\cite{okugawa}
and chiral modes along a screw dislocation.~\cite{imura,sumiyoshi}
It is convenient to modify the tight-binding Hamiltonian
by taking the continuum limit in the $x$ and $y$ directions,
leaving the lattice structure in the $z$ direction
so that the resulting model has a layered structure.
After the partial Fourier transformation in the $z$ direction,
the Hamiltonian is reduced to
\begin{align}
   \mathcal{H} 
     = \left[ 
         \begin{array}{cc}
           \tilde{\Lambda} + 2\gamma\sin(k_{z}a)
              & \tilde{A}(\hat{k}_{x}-i\hat{k}_{y}) \\
           \tilde{A}(\hat{k}_{x}+i\hat{k}_{y})
              & -\tilde{\Lambda} + 2\gamma\sin(k_{z}a)
         \end{array}
       \right] ,
\end{align}
where
$\tilde{\Lambda} = \Delta(k_{z})+\tilde{B}(\hat{k}_{x}^{2}+\hat{k}_{y}^{2})$
with $\hat{k}_{x}=-i\partial_{x}$, $\hat{k}_{y}=-i\partial_{y}$,
$\tilde{A} = Aa$, and $\tilde{B} = Ba^{2}$.
We adapt this model to a cylindrical Weyl semimetal of radius $R$
by using the cylindrical coordinates $(r,\phi)$
with $r = \sqrt{x^{2}+y^{2}}$ and $\phi = \arctan(y/x)$.
Let $\Psi(r,\phi) = {}^t\!(F,G)$ be an eigenfunction of $\mathcal{H}$
for a given $k_{z}$.
It is convenient to rewrite $F$ and $G$ as $F = e^{i\lambda\phi} f(r)$ and
$G = e^{i(\lambda+1)\phi} g(r)$,
where $\lambda$ is the azimuthal quantum number.
Then, in terms of $\psi(r,\phi) = {}^t\!(f,g)$ for given $k_{z}$ and $\lambda$,
the eigenvalue equation is written as
\begin{align}
       \label{eq:ev-Eq1}
   \left[ 
      \begin{array}{cc}
        \Delta(k_{z})-\tilde{B}\mathcal{D}_{\lambda}
           & \tilde{A}\left(-i\partial_{r}-i\frac{\lambda+1}{r}\right) \\
        \tilde{A}\left(-i\partial_{r}+i\frac{\lambda}{r}\right)
           & -\Delta(k_{z})+\tilde{B}\mathcal{D}_{\lambda+1}
      \end{array}
   \right]\psi
   = \tilde{E}\psi ,
\end{align}
where $\tilde{E} = E -2\gamma\sin(k_{z}a)$ and
\begin{align}
      \label{eq:def-D}
 \mathcal{D}_{\lambda} = \partial_{r}^{2}+\frac{1}{r}\partial_{r}
                         - \frac{\lambda^{2}}{r^{2}} .
\end{align}

As demonstrated in Ref.~\citen{takane1},
if $\tilde{B}$ is finite but very small,
the eigenvalue equation, Eq.~(\ref{eq:ev-Eq1}), can be decomposed into
two separate equations: the Weyl and supplementary equations.
The Weyl equation for $f$ and $g$ is given by
\begin{align}
        \label{eq:f-first}
 \left(\mathcal{D}_{\lambda}-\Lambda_{-}\right)f = 0 ,
        \hspace{3mm}
 \left(\mathcal{D}_{\lambda+1}-\Lambda_{-}\right)g = 0 ,
\end{align}
while the supplementary equation is
\begin{align}
        \label{eq:f-second}
 \left(\mathcal{D}_{\lambda}-\Lambda_{+}\right)f = 0 ,
        \hspace{3mm}
 \left(\mathcal{D}_{\lambda+1}-\Lambda_{+}\right)g = 0 ,
\end{align}
where
\begin{align}
          \label{eq:def-Lambda}
  \Lambda_{-} = -\frac{\tilde{E}^{2}-\Delta^{2}}{\tilde{A}^{2}} ,
         \hspace{6mm}
  \Lambda_{+} = \frac{\tilde{A}^{2}}{\tilde{B}^{2}} .
\end{align}
Here, $f$ and $g$ are related by the original eigenvalue equation
with a finite but very small $\tilde{B}$.
Note that the restriction on $\tilde{B}$ (i.e., $\tilde{B}$ is very small)
does not significantly affect the behaviors of chiral surface states.
Indeed, the energy of chiral surface states does not depend on $\tilde{B}$
as seen in Eq.~(\ref{eq:disp-CS_state}).

We hereafter focus on chiral surface states,
which appear only in the case of $|\Delta(k_{z})|>|\tilde{E}|$.
The solutions of both the Weyl and supplementary equations are expressed
by modified Bessel functions in this case.
By superposing two solutions
that asymptotically increase in an exponential manner,
we can describe spatially localized states near the side.
With $\eta \equiv \sqrt{\Delta^{2}-\tilde{E}^{2}}/\tilde{A}$
and $\kappa \equiv \tilde{A}/\tilde{B}$,
the general solution for given $\lambda$ and $k_{z}$ is written as
\begin{align}
  \psi = a \left[
              \begin{array}{c}
                 I_{|\lambda|}(\eta r) \\
                 -i\frac{\Delta-\tilde{E}}{\sqrt{\Delta^{2}-\tilde{E}^{2}}}
                 I_{|\lambda+1|}(\eta r)
              \end{array}
           \right]
       + b \left[
              \begin{array}{c}
                 I_{|\lambda|}(\kappa r) \\
                 iI_{|\lambda+1|}(\kappa r)
              \end{array}
           \right] ,
\end{align}
where the first and second terms respectively arise from
the Weyl and supplementary equations.
The boundary condition of $\psi(R)={}^t\!(0,0)$ requires
\begin{align}
  \frac{\Delta-\tilde{E}}{\sqrt{\Delta^{2}-\tilde{E}^{2}}}
  = -\frac{I_{|\lambda+1|}(\kappa R)}{I_{|\lambda|}(\kappa R)}
    \frac{I_{|\lambda|}(\eta R)}{I_{|\lambda+1|}(\eta R)} ,
\end{align}
indicating that a relevant solution is obtained only in the case of
$\Delta(k_{z}) < 0$, which holds when $k_{z} \in (-k_{0}, k_{0})$.
The eigenvalue of energy is approximately determined as
\begin{align}
       \label{eq:disp-CS_state}
  E = \frac{\tilde{A}}{R}\left(\lambda+\frac{1}{2}\right)+2\gamma\sin(k_{z}a) .
\end{align}
In the case of $\gamma = 0$, the dispersion is flat (i.e., independent of
$k_{z}$), representing a characteristic feature of the chiral surface state.
An electron in the chiral surface state propagates
in the anticlockwise direction viewed from above.
The dispersion becomes dependent on $k_{z}$ if $\gamma \neq 0$,
indicating that the group velocity is tilted to the $z$ direction.
Consequently, an electron in the chiral surface state circulates around
the side surface in a spiral manner.~\cite{baireuther}
This implies that a spontaneous charge current in the $z$ direction
can be induced near the side surface if $\gamma \neq 0$.

Now, we roughly determine the magnitude of a spontaneous circulating current
in the system consisting of $N$ layers.
The circulating charge current carried by each chiral surface state is
\begin{align}
   J_{\phi}^{0} = -e\frac{\tilde{A}}{2\pi R} ,
\end{align}
which flows in the clockwise direction.
Note that the chiral surface state with $E(\lambda)$ appears
only when $k_{z} \in (-k_{0},k_{0})$.
If $k_{z}$ deviates from this interval, the state is continuously transformed
to a bulk state, which is spatially extended over the entire system.
Let us determine the total charge current $J_{\phi}$.
Since $J_{\phi}$ vanishes at $E_{F} = 0$, we need to collect
the contributions to $J_{\phi}$ arising from the chiral surface states
with $E(\lambda)$ satisfying $0 < E < E_{F}$ if $E_{F} > 0$.
Here, it is assumed that the contribution from bulk states in the same interval
of energy is not important
since their wavelength is relatively long and hence
they cannot induce a short-wavelength response localized near the surface.
If $E_{F} < 0$, the total charge current is obtained by collecting
the contributions to $J_{\phi}$ arising from the states
with $E(\lambda)$ satisfying $E_{F} < E < 0$ and then reversing its sign.
In addition to the condition for $\lambda$, it is important to note that
the chiral surface states are stabilized only when $|k_{z}| < k_{0}$.
Assuming that $k_{z}$ is given by $k_{z}^{j} = j\pi/[(N+1)a]$
with $j = 1,2,3,\dots$ in the system consisting of $N$ layers,
we require $k_{z}^{j} < k_{0}$.
From the observation given above, we find that
the spontaneous charge current per layer is
\begin{align}
      \label{eq:J_phi-av}
   \frac{J_{\phi}}{N} = -e\frac{k_{0}a}{2\pi^{2}} E_{F} ,
\end{align}
which depends on $E_{F}$ as well as $k_{0}$.
This indicates that the direction of flow
reverses depending on the sign of $E_{F}$.
That is, the spontaneous current flows in opposite directions
for the cases of electron doping and hole doping.
Note that, since $J_{\phi}/N$ is independent of $R$, we expect that
the circulating charge current will be insensitive to the geometry of
the Weyl semimetal.
Thus, we expect that Eq.~(\ref{eq:J_phi-av}) can be applied to the system of
a rectangular parallelepiped, which is treated in the next section.

%%%%%%%%%%%%%%%%%%
\begin{figure}[btp]
\begin{center}
\includegraphics[width=6cm]{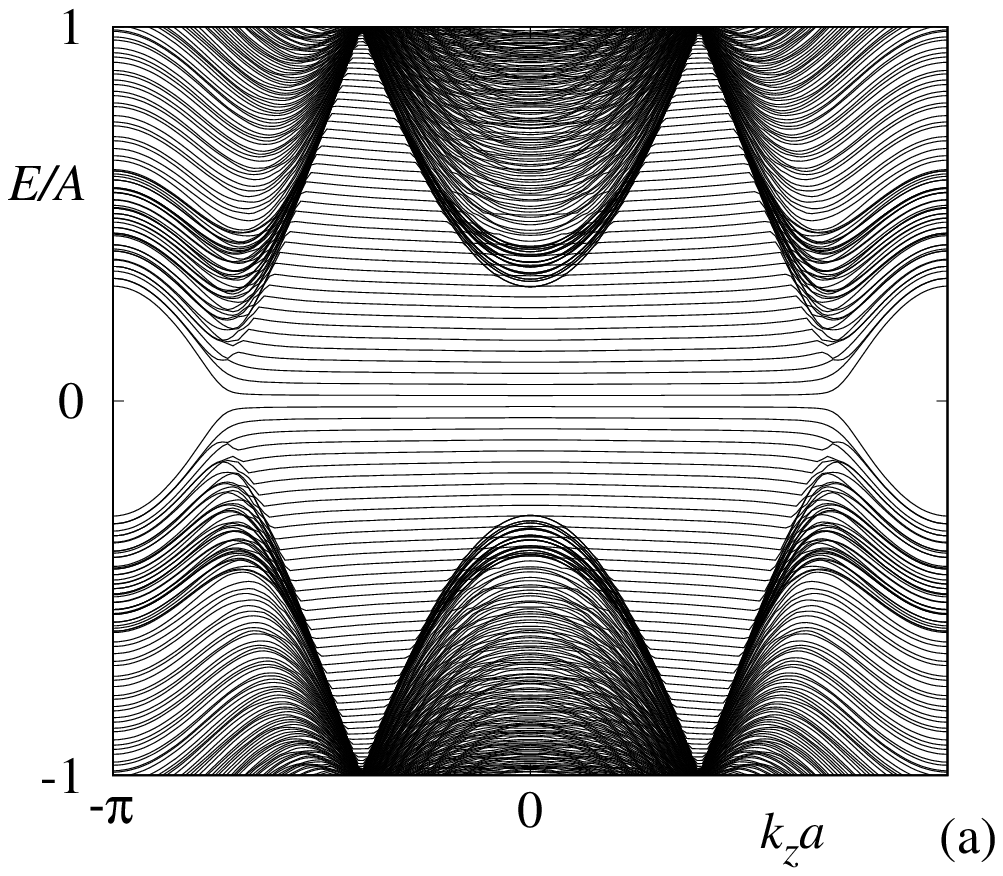}
\includegraphics[width=6cm]{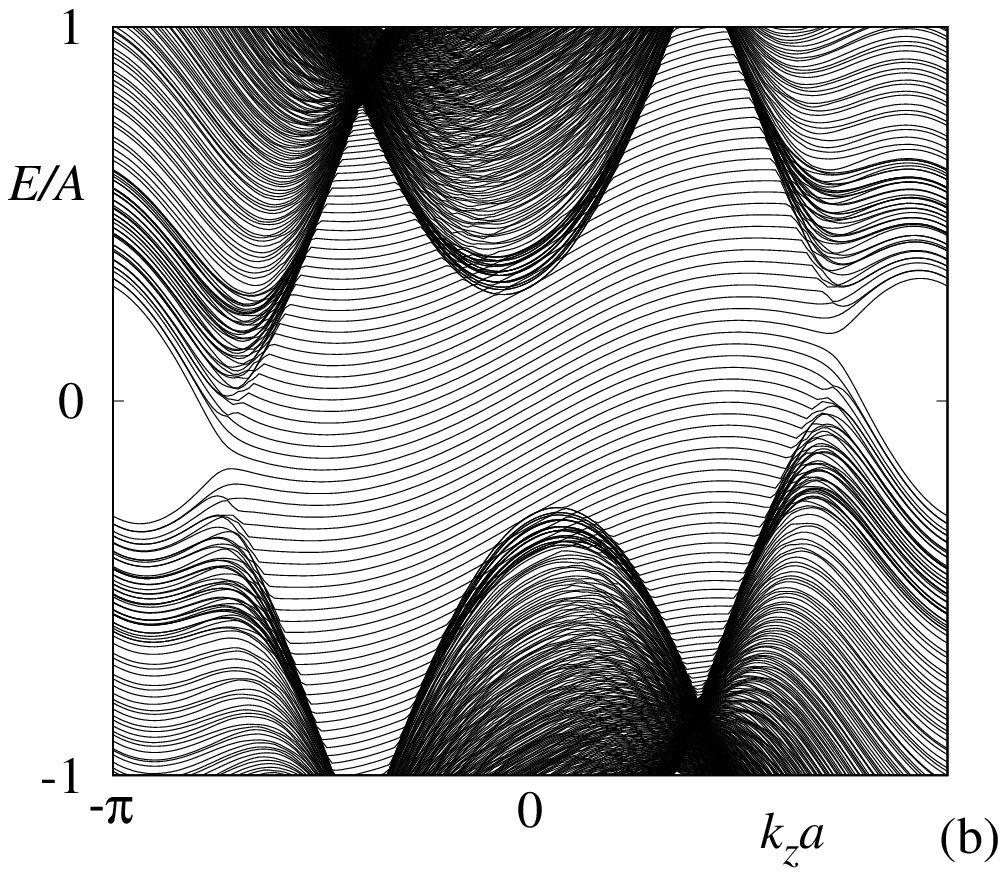}
\end{center}
\caption{
Energy dispersion as a function of $k_{z}$ in the cases of
(a) $\gamma/A =0$ and (b) $0.1$.
For the purely bulk states,
only a quarter of the corresponding branches are shown for clarity.
}
\end{figure}
%%%%%%%%%%%%%%%%%%

\section{Numerical Results}

We focus on the lattice system of the rectangular parallelepiped shown in
Fig.~1, which occupies the region of $1 \le l, m \le L$ and $1 \le n \le N$,
under the open boundary condition in the three spatial directions.
We set $L = 50$ and $N = 30$ with the following parameters:
$B/A =0.5$, $t/A = 0.5$, and $k_{0}a = 3\pi/4$.
We consider the cases of $\gamma/A = 0$ and $0.1$
for $E_{F}/A = \pm 0.1$ and $\pm 0.2$.
The energy dispersion for the infinitely long system
with a cross-sectional area of $L^{2}$ is shown in Fig.~2 in the cases of
(a) $\gamma/A = 0$ and (b) $0.1$ as a function of $k_{z}$.
In the case of $\gamma/A = 0$, branches with flat dispersion are uniformly
distributed near $E = 0$ in the region of $-k_{0} < k_{z} < k_{0}$.
They represent chiral surface states localized near
the side surface of the system.
These states should induce a circulating charge current near the side surface
when $E_{F} \neq 0$.
The dispersion of these states becomes slightly upward to the right
in the case of $\gamma/A = 0.1$, implying the appearance of
a spontaneous charge current in the $z$ direction.
Hereafter, we numerically study how the spontaneous charge current appears
in this system depending on $E_{F}$ and $\gamma$.

%%%%%%%%%%%%%%%%%%
\begin{figure}[btp]
\begin{center}
\includegraphics[width=6.0cm]{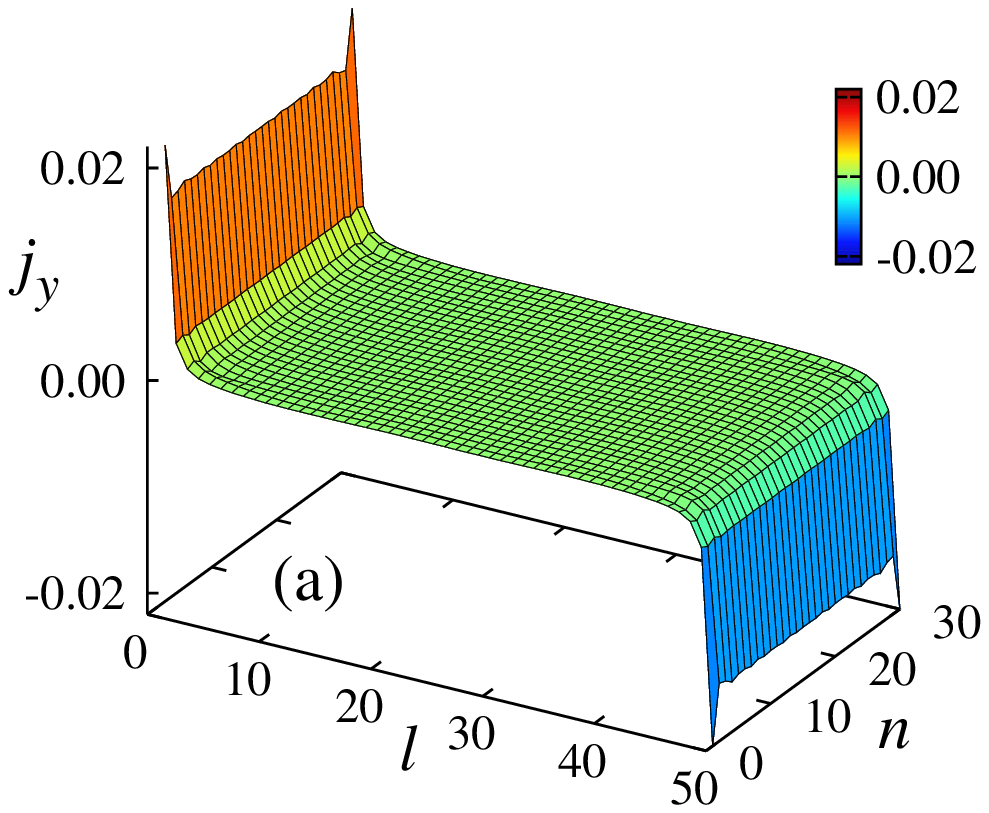}
\includegraphics[width=6.0cm]{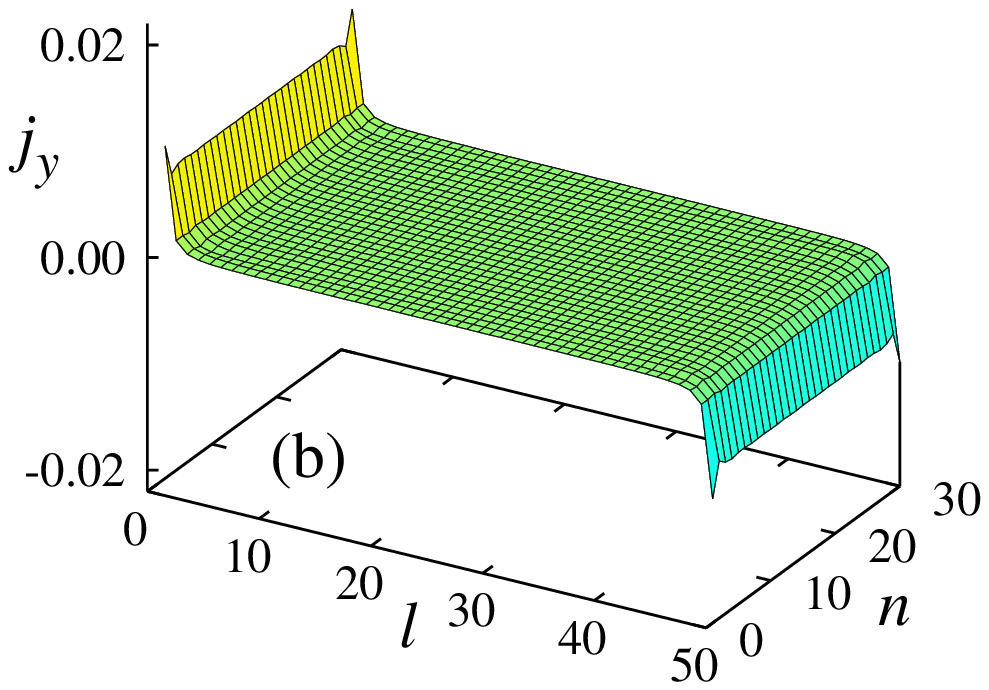}
\includegraphics[width=6.0cm]{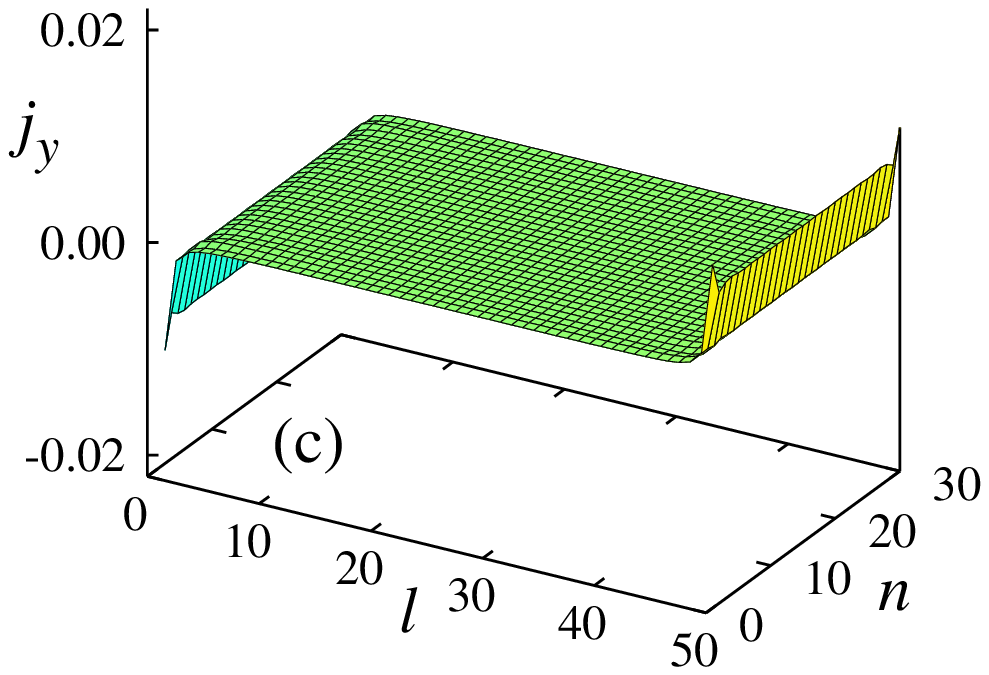}
\includegraphics[width=6.0cm]{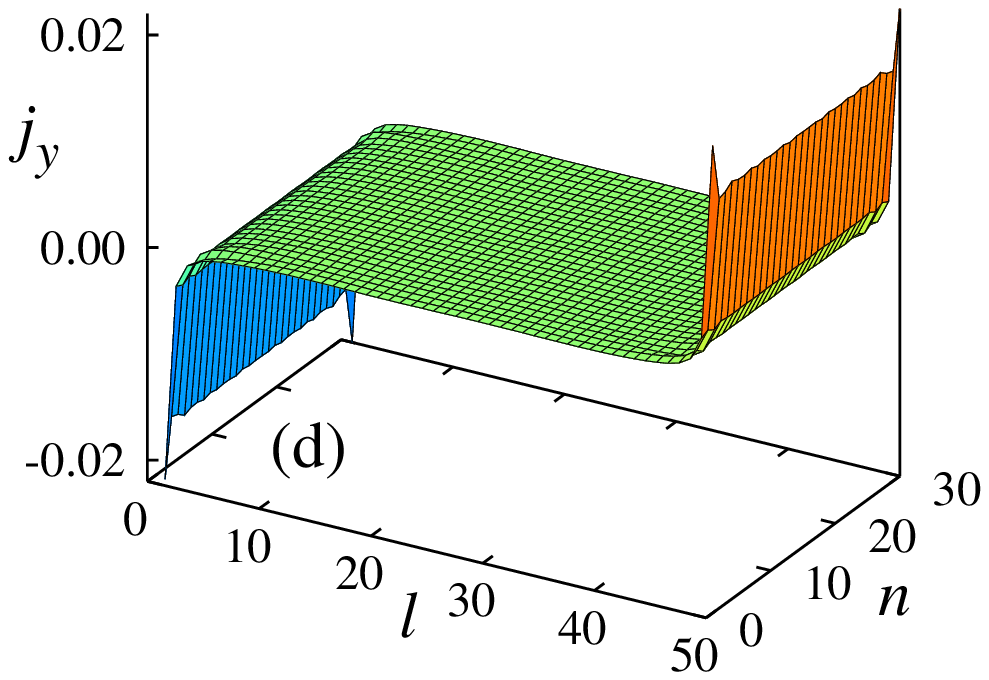}
\end{center}
\caption{(Color online)
Spatial distribution of $j_{y}$ normalized by $eA$ in the cross section
parallel to the $xz$ plane; (a) $E_{F}/A = 0.2$, (b) $0.1$, (c) $-0.1$,
and (d) $-0.2$.
}
\end{figure}
%%%%%%%%%%%%%%%%%%
Firstly, let us examine the behaviors of a spontaneous charge current
that circulates around the system near the side surface.
To do so, we calculate the distribution of the spontaneous charge current
in the $y$ direction through the cross section parallel to the $xz$ plane
at the center of the system (dotted line in Fig.~1).
Precisely speaking, $j_{y}$ on each link connecting
the $(l,25,n)$th and $(l,26,n)$th sites is calculated
for $1 \le l \le 50$ and $1 \le n \le 30$.
Figure~3 shows the results for $j_{y}$ normalized by $eA$ for $\gamma/A = 0$
and $E_{F}/A = 0.2$, $0.1$, $-0.1$, and $-0.2$.
The results for $\gamma/A = 0.1$ are not shown
since they are almost identical to those for $\gamma/A = 0$.
We observe that $j_{y} > 0$ near $l = 1$ and $j_{y} < 0$ near $l = 50$
in the case of $E_{F} > 0$, whereas $j_{y} < 0$ near $l = 1$
and $j_{y} > 0$ near $l = 50$ in the case of $E_{F} < 0$.
This indicates that the spontaneous charge current circulates around the system
in the clockwise direction viewed from above when $E_{F} > 0$,
whereas it circulates in the anticlockwise direction
when $E_{F} < 0$ (see Fig.~1).
That is, its direction of flow is opposite for
the cases of $E_{F} > 0$ and $E_{F}<0$.
We also observe that $j_{y}$ increases
with increasing $E_{F}$ in accordance with Eq.~(\ref{eq:J_phi-av}).
Equation~(\ref{eq:J_phi-av}) predicts $|J_{y}|/N \approx 0.024 \times eA$
in the case of $E_{F}/A = \pm 0.2$, where $J_{y}$ represents
the total charge current induced near each side surface of height $N$.
This result is consistent with those shown in Figs.~3(a) and 3(d).

%%%%%%%%%%%%%%%%%%
\begin{figure}[btp]
\begin{center}
\includegraphics[width=6.0cm]{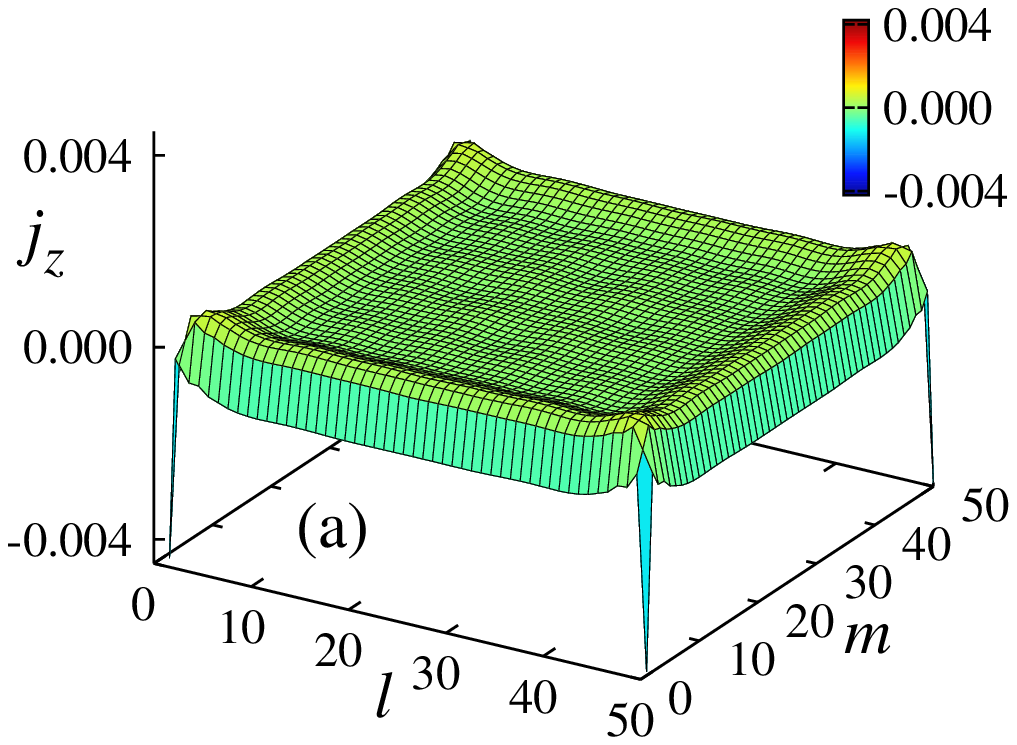}
\includegraphics[width=6.0cm]{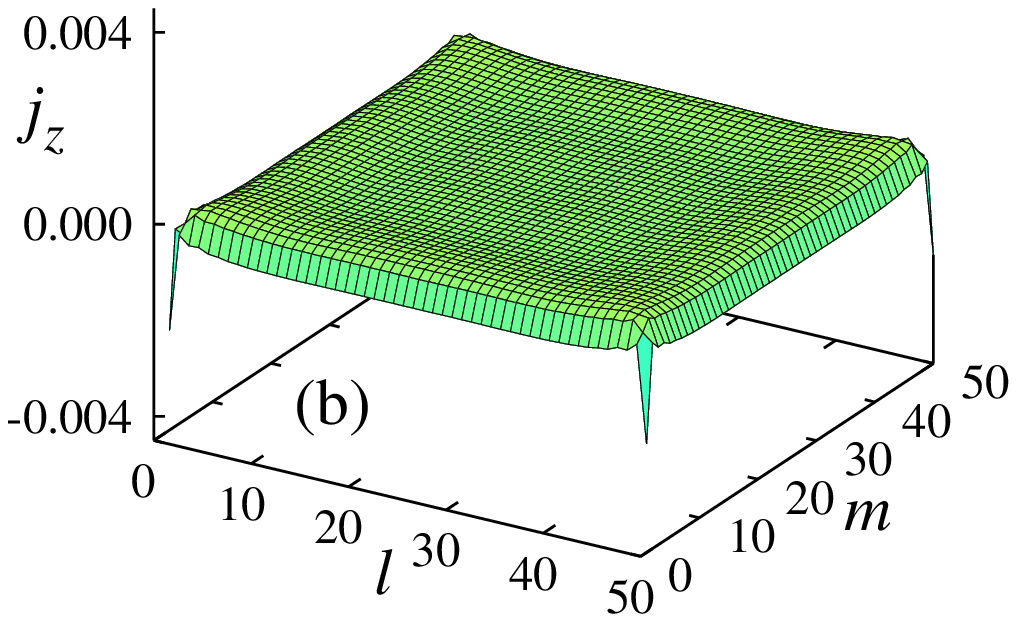}
\includegraphics[width=6.0cm]{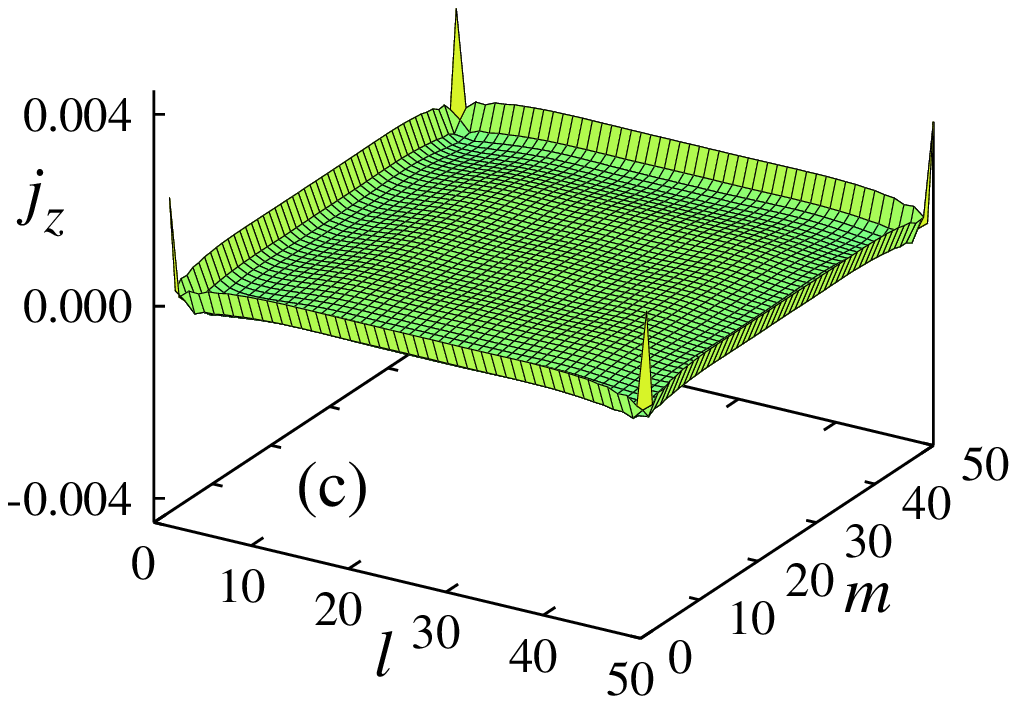}
\includegraphics[width=6.0cm]{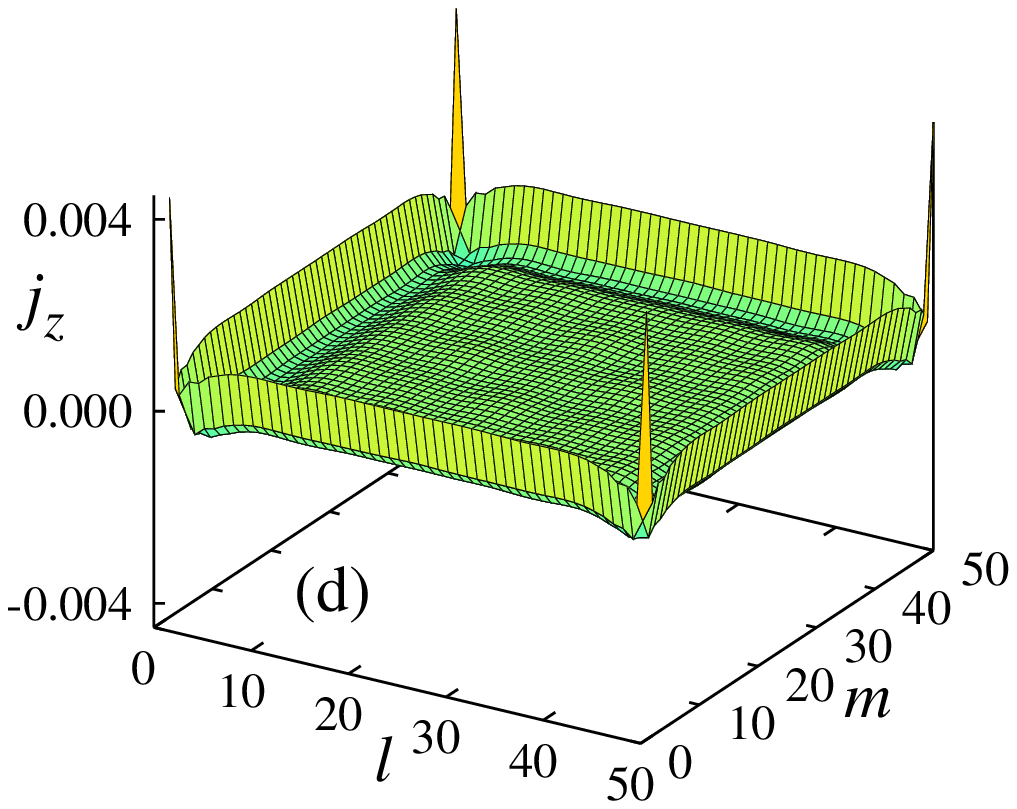}
\end{center}
\caption{(Color online)
Spatial distribution of $j_{z}$ normalized by $eA$ in the cross section
parallel to the $xy$ plane; (a) $E_{F}/A = 0.2$, (b) $0.1$, (c) $-0.1$,
and (d) $-0.2$.
}
\end{figure}
%%%%%%%%%%%%%%%%%%
Secondly, let us examine the behaviors of a spontaneous charge current
in the longitudinal direction.
We calculate the distribution of the charge current in the $z$ direction
through the cross section parallel to the $xy$ plane
at the center of the system (broken line in Fig.~1).
Precisely speaking, $j_{z}$ on each link connecting
the $(l,m,15)$th and $(l,m,16)$th sites is calculated
for $1 \le l \le 50$ and $1 \le m \le 50$.
Figure~4 shows the results for $j_{z}$ normalized by $eA$ for $\gamma/A = 0.1$
and $E_{F}/A = 0.2$, $0.1$, $-0.1$, and $-0.2$.
The results for $\gamma/A = 0$ are not shown
since $j_{z}$ vanishes everywhere in this case.
Again, Fig.~4 indicates that $j_{z}$ increases with increasing $E_{F}$
and that its sign is opposite for the cases of $E_{F} > 0$ and $E_{F}<0$.
Note that a relatively large current appears near the side surface,
particularly near the corners, while a small current flowing
in the opposite direction is distributed beneath the side surface.
The former is induced by chiral surface states,
while the latter originates from bulk states.
These two contributions cancel each other out
if they are integrated over the cross section;
thus, the total charge current in the $z$ direction completely vanishes.

%%%%%%%%%%%%%%%%%%
\begin{figure}[btp]
\begin{center}
\includegraphics[width=6.0cm]{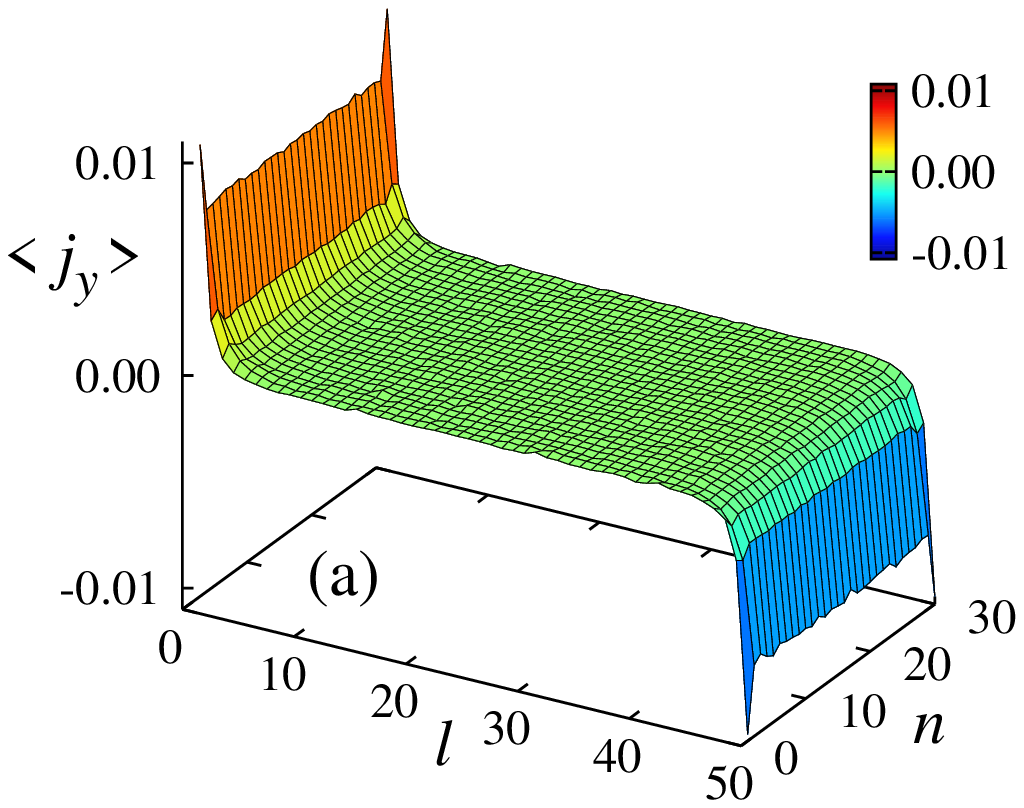}
\includegraphics[width=6.0cm]{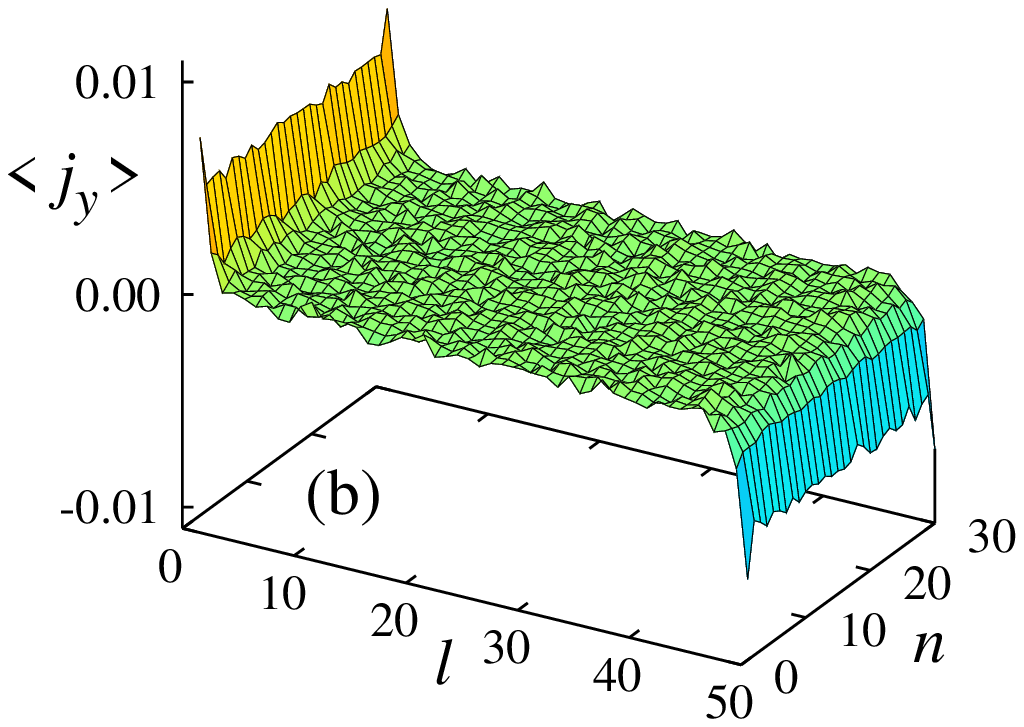}
\includegraphics[width=6.0cm]{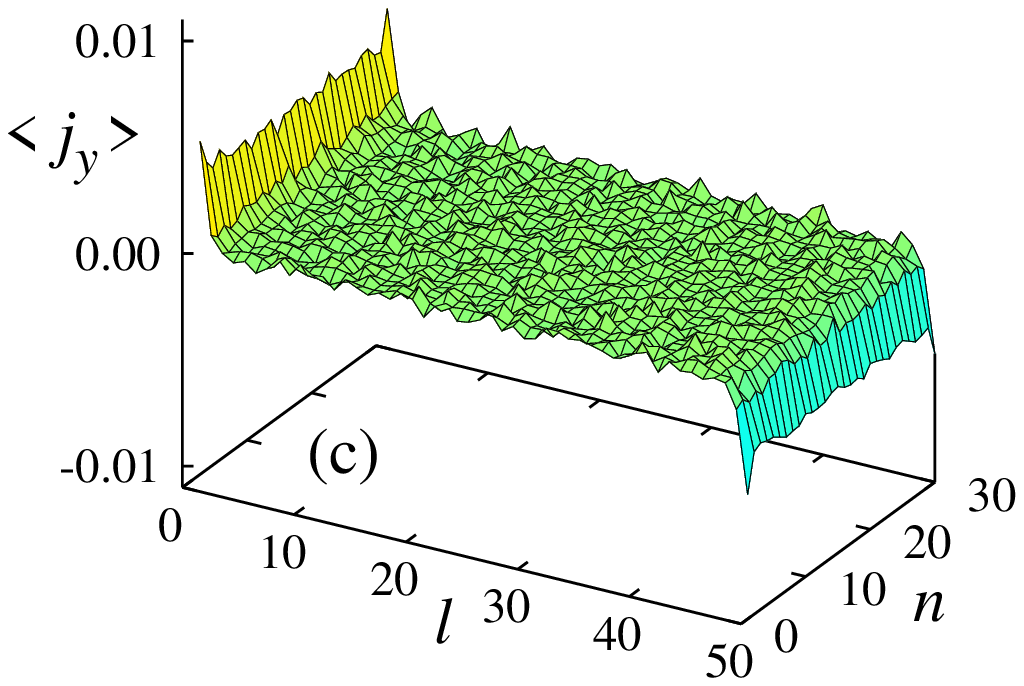}
\end{center}
\caption{(Color online)
Spatial distribution of $\langle j_{y}\rangle$ normalized by $eA$
in the cross section parallel to the $xz$ plane at $E_{F}/A = 0.1$;
(a) $W/A = 3$, (b) $4$, and (c) $4.5$.
}
\end{figure}
%%%%%%%%%%%%%%%%%%
%%%%%%%%%%%%%%%%%%
\begin{figure}[btp]
\begin{center}
\includegraphics[width=6.0cm]{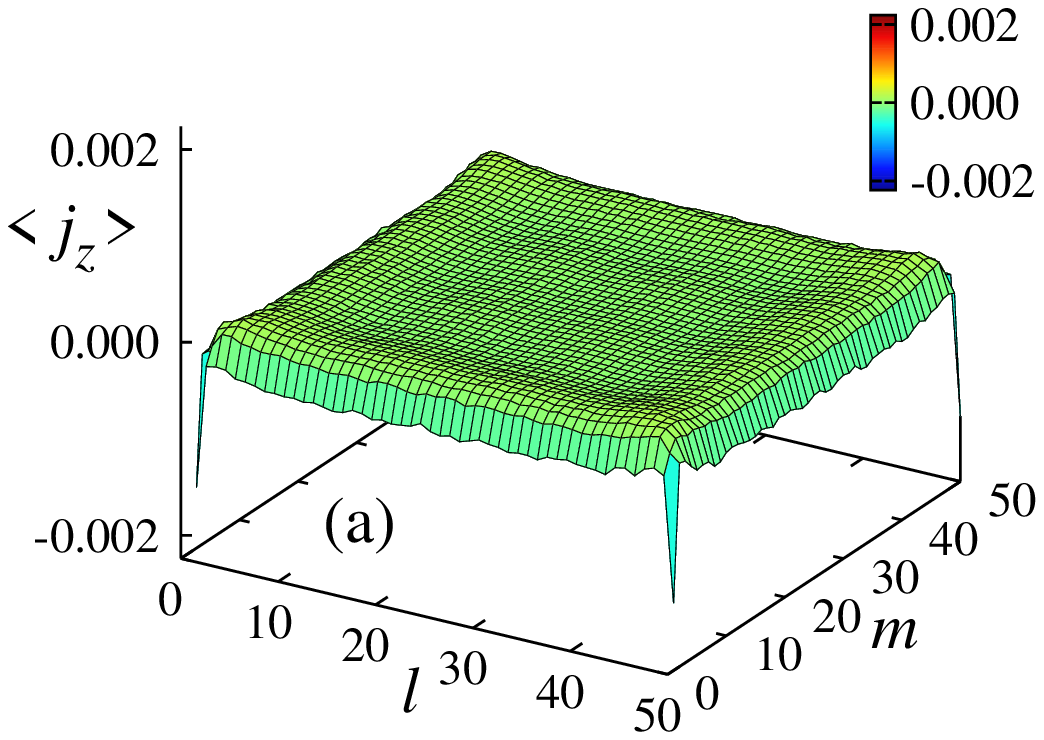}
\includegraphics[width=6.0cm]{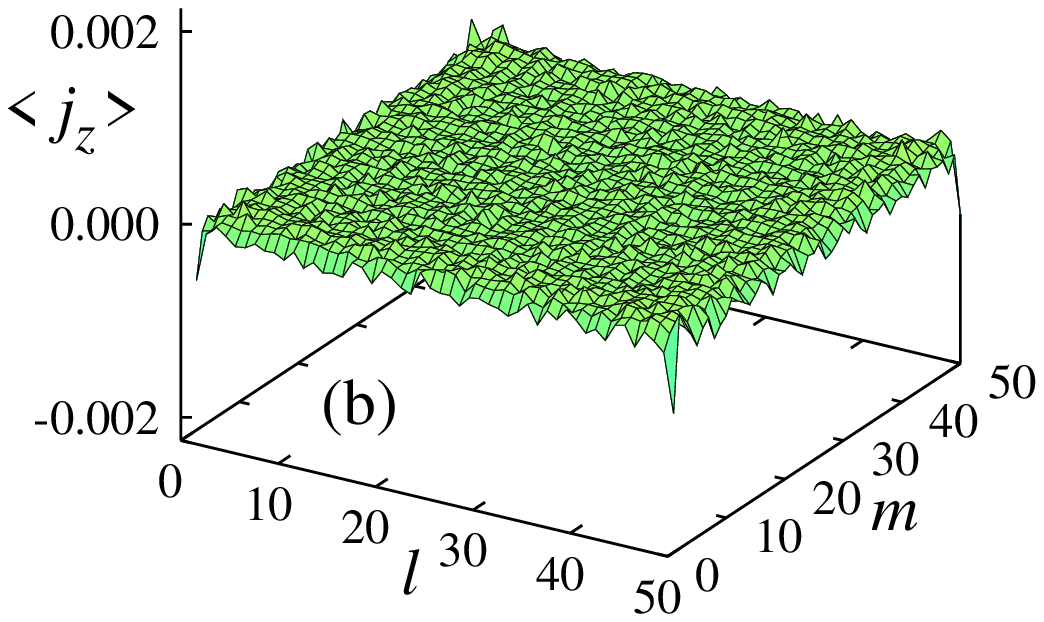}
\end{center}
\caption{(Color online)
Spatial distribution of $\langle j_{z}\rangle$ normalized by $eA$
in the cross section parallel to the $xy$ plane at $E_{F}/A = 0.1$;
(a) $W/A = 2$, and (b) $3$.
}
\end{figure}
%%%%%%%%%%%%%%%%%%
Finally, we examine the effect of disorder~\cite{chen2,shapourian,LOS,gorbar,
takane2,yoshimura} on the spontaneous charge current
by adding the impurity potential term
\begin{align}
  H_{\rm imp}
  = \sum_{l,m,n} |l,m,n \rangle
                   \left[ \begin{array}{cc}
                             V_{1}^{(l,m,n)} & 0 \\
                             0 & V_{2}^{(l,m,n)}
                          \end{array} \right]
                 \langle l,m,n|
\end{align}
to the Hamiltonian $H$, where $V_{1}$ and $V_{2}$ are assumed to be
uniformly distributed within the interval of $[-W/2,+W/2]$.
Previous studies have shown that a Weyl semimetal phase is robust against
weak disorder up to a critical disorder strength,
$W_{\rm c}$,~\cite{chen2,shapourian} and that chiral surface states also
persist as long as $W < W_{\rm c}$.~\cite{takane2}
In the case of $B/A =0.5$, $t/A = 0.5$, and $k_{0}a = 3\pi/4$,
the critical disorder strength is $W_{\rm c}/A \sim 4$.
The ensemble averages, $\langle j_{y}\rangle$ and $\langle j_{z}\rangle$,
are calculated over $500$ samples with different impurity configurations
at $E_{F}/A = 0.1$ for a given value of $W/A$.
In calculating $j_{y}$ and $j_{z}$ for a given impurity configuration,
we take account of only the contribution from electron states
with an energy $E$ satisfying $0 < E < E_{F}$,
assuming that electron states below the band center
have no contribution owing to cancellation between them.
This assumption is not strictly justified here
since particle-hole symmetry is broken by the impurity potential.
Nonetheless, this should be a good approximation
after taking the ensemble average.

Figure~5 shows the results for $\langle j_{y}\rangle$ normalized by $eA$
for $\gamma/A = 0$ and $E_{F}/A = 0.1$ with $W/A = 3$, $4$, and $4.5$.
We observe that the circulating charge current is robust against disorder
up to $W/A \sim 4$ but is suppressed when $W/A$ exceeds $4$.
This behavior is consistent with
an observation reported previously.~\cite{takane2}
Figure~6 shows the results for $\langle j_{z}\rangle$ normalized by $eA$
for $\gamma/A = 0.1$ and $E_{F}/A = 0.1$ with $W/A = 2$ and $3$.
We observe that $\langle j_{z}\rangle$ is significantly suppressed
in the case of $W/A = 3$,
although the circulating charge current is almost unaffected in this case.
This indicates that the charge current in the $z$ direction is more fragile
than the circulating charge current against the mixing of
chiral surface states and bulk states due to disorder.

\section{Summary and Discussion}

We theoretically studied a spontaneous charge current
due to chiral surface states in the ground state of a Weyl semimetal.
We analytically and numerically determined the magnitude of
the charge current induced near the side surface of the system.
It is shown that no spontaneous charge current appears
when the Fermi level, $E_{F}$, is located at the band center.
It is also shown that, once $E_{F}$ deviates from the band center,
the spontaneous charge current appears to circulate around the side surface
of the system and its direction of flow is opposite for the cases of
electron doping and hole doping.
The circulating current is shown to be robust against weak disorder.

Let us focus on the two features revealed in this paper: the appearance of
a spontaneous charge current except at the band center
and the reversal of its direction of flow as a function of $E_{F}$.
As they are derived by using a model possessing particle-hole symmetry,
a natural question arises: do these features manifest themselves
even in the absence of particle-hole symmetry?
The answer is yes.
The disappearance of the spontaneous charge current reflects the fact that
the contribution from chiral surface states is
completely canceled out by that from bulk states.
As the spontaneous charge current due to chiral surface states
is localized near the side surface, this cancellation should be
mainly caused by bulk states with a short wavelength,
occupying the bottom region of the energy band far from the band center.
Hence, if $E_{F}$ is varied near the band center, the contribution from
bulk states is almost unaffected but that from chiral surface states
is significantly changed, depending on $E_{F}$ in a roughly linear manner.
This behavior should take place
regardless of the presence or absence of particle-hole symmetry.
Thus, we expect that the features of the spontaneous charge current
still manifest themselves even in the absence of particle-hole symmetry,
although the point of the disappearance shifts away from the band center.

\section*{Acknowledgment}

This work was supported by JSPS KAKENHI Grant Numbers
JP15K05130 and JP18K03460.

\end{document}